\documentclass[dvips]{acta}
\usepackage{supertabular,lscape,epsfig}
\usepackage{amssymb}
\usepackage{amsmath}
\usepackage{txfonts}

\SetPages{0}{0}

\SetVol{00}{2014}

\begin{document}

\begin{Titlepage}

\Title{The study of triple systems V949~Cen, V358~Pup, and V1055~Sco\footnote[1]{Based on
observations made with ESO Telescopes at the La Silla Paranal Observatory under programmes ID
082.D-0499, 084.D-0591, 085.C-0614, 086.D-0078, and 190.D-0237.}} \Author{Zasche, P.$^{1}$,
Ho\v{n}kov\'a, K.$^{2}$, Jury\v{s}ek, J.$^{1,3}$, Ma\v{s}ek, M.$^{3}$}
 {$^{1}$ Astronomical Institute, Charles University Prague, Faculty of Mathematics and Physics,
 V Hole\v{s}ovi\v{c}k\'ach 2, Praha 8, CZ-180 00, Czech Republic\\
e-mail: zasche@sirrah.troja.mff.cuni.cz \\
 $^{2}$ Variable Star and Exoplanet Section of Czech Astronomical Society\\
 $^{3}$ Institute of Physics, Czech Academy of Sciences, 182 21 Praha, Czech Republic}

\Received{Month Day, Year}
\end{Titlepage}

\Abstract{The systems V949~Cen, V358~Pup, and V1055~Sco are triples comprised of an eclipsing
 binary orbiting with a distant visual component on a much longer orbit. The first detailed
photometric analysis of these interesting systems was performed using also the archival data from
Hipparcos, ASAS, SuperWASP, OMC, and Pi Of The Sky surveys. The system V358~Pup was also analysed
using the archival ESO spectra and the radial velocities were derived. The analyses of their light
curves revealed the physical properties of the eclipsing components, while the interferometric data
for these systems obtained during the last century show that the binaries are also weakly
gravitationally bounded with the third components on much longer orbits. The photometry was carried
out with the robotic telescope FRAM (part of the Pierre Auger Cosmic Ray Observatory), located in
Argentina. The $BVRI$ light curves were analysed with the {\sc PHOEBE} program, yielding the basic
physical parameters of the systems and their orbits. V949~Cen and V358~Pup were found to be
detached systems, while V1055~Sco is probably a semi-detached one. V358~Pup shows a slow apsidal
motion, while for V1055~Sco we detected some period variation probably due to the third body in the
system, which cannot easily be attributed to the close visual companion. Therefore, we speculate
that V1055~Sco can be a quadruple system. For V949~Cen a new orbit was computed, having the orbital
period of about 855~yr. }

 {binaries: eclipsing -- binaries: visual -- stars: fundamental parameters -- stars: individual:
V949~Cen, V358~Pup, V1055~Sco.}

\section{Introduction}

The eclipsing binaries as parts of the multiple stellar systems (i.e. of multiplicity three and
higher) are excellent objects to be studied. The importance of such systems lies in the fact that
one can study the stellar evolution in them, their origin, tidal interaction between the
components, testing the influence of the distant companions to the close pair, a role of Kozai
cycles, studying the dynamical effects and precession of the orbits, but also the statistics and
relative frequency of such systems among the stars in our Galaxy (and outside), see e.g. Tokovinin
(2007), Guinan \& Engle (2006), or Goodwin \& Kroupa (2005). On the other hand, there are still
plenty of interesting multiple systems, which were not analysed yet.

\section{Photometric observations}

All of the new photometric observations were obtained in Argentina, using the FRAM telescope
(Prouza et al. 2010). It is located near the Malarg\"ue town, about 1400 m above a sea level. The
instrument used for observing V949~Cen and V358~Pup was a classical telescope of Schmidt-Cassegrain
type, 0.3~m in diameter, equipped with a G2-1600 CCD camera. On the other hand, for V1055~Sco only
a small Nikkor lens with 107~mm diameter and a CCD camera of G4-16000 type was used. Both
instruments are equipped with standard Johnson $BVRI$ filters, see Bessell (1990). All frames have
been reduced by differential aperture photometry using software package
C-MuniPack\footnote{http://c-munipack.sourceforge.net/}, the adaptation of the MuniPack code (Hroch
1998), based on the DaoPhot routines (Stetson 1987). The G2-1600 CCD camera sensor area is of
1536$\times$1024 square pixels (pixel linear dimension is 9~$\mu$m), which leads to the field of
view of 24$^\prime$$\times$16$^\prime$ (0.93 arcsec/px), while the second CCD camera G4-16000 has
the sensor area of 4094$\times$4094 square pixels (pixel linear dimension is 9~$\mu$m), which leads
to the field of view of 7$^\circ$$\times$7$^\circ$ (6.19 arcsec/px). The radii of the apertures for
our studied stars were 7.09~px for V1055~Sco, 7.09~px for V358~Pup, and 11.82~px for V949~Cen,
respectively. The star HD~113916 was used as a comparison for V949~Cen, TYC~7636-2147-1 was used
for V358~Pup, while HD~147704 was used for V1055~Sco.

The older archival data were taken from the publicly available databases such as Hipparcos, ASAS,
SuperWASP, OMC, and Pi Of The Sky surveys. However, these photometric observations were used mainly
for the derivation of the minima times for a subsequent period analysis. For the {\sc PHOEBE} light
curve fitting the individual filter passbands used were: $V$ for ASAS, $H_p$ for Hipparcos, and
$H_p$ also used for SuperWASP due to unavailable SuperWASP filter in {\sc PHOEBE} and relative
similarity of transmission curves of $H_p$ and SuperWASP filters.

\section{Analysis}

After the reduction of the photometric frames, the FRAM photometry was used for a subsequent
analysis of the light curves. Program {\sc PHOEBE} ver. 0.31 (Pr\v{s}a \& Zwitter 2005) was used,
which is based on a standard Wilson-Devinney code (Wilson \& Devinney 1971) and its later
modifications. Because of having only limited information about the systems, several parameters of
the light curve model have to be fixed for the modelling. For the systems with no radial
velocities, the individual masses of the components are not known. Hence, the mass ratio was kept
fixed at a value of  $q=1.0$ for the detached systems (see below). The primary temperature was
fixed at a value derived from the photometric indices, or according to the published spectral type.
The albedo and gravity darkening coefficients were kept fixed. On the other hand, the value of the
third light was also computed due to the fact that all of these systems are triple stars.

\section{Individual systems}

For our analysis, we have selected three interesting systems. These systems were chosen according
to the following selection criteria. All of them are relatively bright ($<$10mag), are on the
southern sky, all have periods shorter than 10 days, all are multiples (eclipsing binary with a
third visual component) and all are also only seldom-investigated systems. Therefore, three stars
were chosen: V949~Cen, V358~Pup, and V1055~Sco. In the following subsections we will focus on the
individual systems in more detail.

\subsection{V949~Cen}

V949~Cen (also HD 113840, HIP 64025 ) is a variable star of Algol type. Its orbital period is of
about 3.8 days, while its brightness in maximum is of about 9~mag in $V$~filter. It was also
discovered as a double star by Finsen (1928), and since then 8 astrometric observations of the
visual pair were carried out, see Table \ref{AstrObsV949Cen}. Houk \& Cowley (1975) derived its
spectral type as G1/G2V. Its parallax is rather uncertain, because the Hipparcos catalogue (van
Leeuwen 2007) gave the value of parallax $1.42 \pm 1.48~$mas.

\begin{figure}
  \centering
  \includegraphics[width=0.75\textwidth]{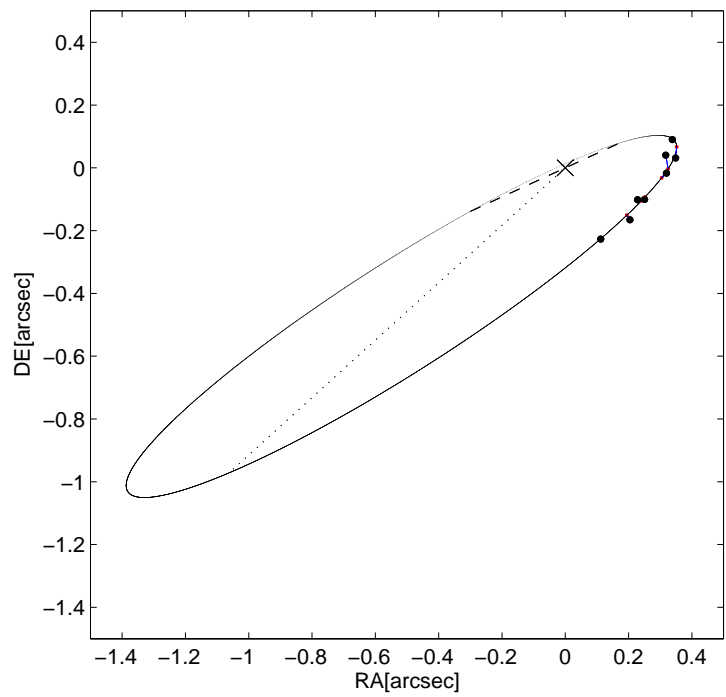}
  \caption{The orbit of the visual pair of V949~Cen. The individual observations are plotted as dots,
  connected with their theoretical positions on the orbit with short abscissae. The dotted
  line represents the line of the apsides, while the dashed one stands for the line of the nodes.
  The eclipsing binary is located in the coordinates (0,0).}
  \label{FigV949CenOrbit}
\end{figure}

\begin{table}
 \begin{minipage}{130mm}
 \centering
  \caption{The individual astrometric observations of V949 Cen.}  \label{AstrObsV949Cen}
  \begin{tabular}{@{}c c c c@{}}
\hline
                &  Pos.Angle &  Separation  & Reference \\
 Year           &  [degree]  &  [arcsec]    &           \\
 \hline
       1927.48   &  284.9   &    0.35    &  Finsen (1928)      \\
       1935.84   &  275.1   &    0.35    &  Rossiter (1955)    \\
       1953.37   &  277.2   &    0.32    &  van den Bos (1956) \\
       1960.49   &  267.0   &    0.32    &  van den Bos (1961) \\
       1976.128  &  248.1   &    0.27    &  Worley (1978)      \\
       1979.19   &  246.0   &    0.25    &  Heintz (1980)      \\
       1991.25   &  231.0   &    0.263   &  Perryman et al. (1997) \\
       2011.0401 &  206.3   &    0.2532  &  Hartkopf et al. (2012) \\
 \hline
\end{tabular}
\end{minipage}
\end{table}

No radial velocity analysis as well as no light curve analysis were published. On the other hand,
from the available astrometric observations a preliminary visual orbit was derived, see Hartkopf \&
Mason (2011). Their analysis resulted in orbital period of about 348~yr, but the individual
observations rather deviate from the fit and the most recent one is very distant from its predicted
position. Hence, we performed an updated orbital fitting, resulting in parameters given in Table
\ref{AstrOrbitV949Cen}. The fit is presented in Fig. \ref{FigV949CenOrbit} and as one can see, it
is still very preliminary yet, because the arc of the orbit covered with the observations is still
rather limited. However, we have to mention that the solution is not self-consistent due to high
value of the total mass from the Kepler's third law (assuming 3 G1V stars, i.e. $3 \cdot 1.15$
M$_\odot = 3.45$ M$_\odot$). Hence, the explanation is that the period $p_3$ should be much longer
(up to the 39000~yr), or other solution is that its distance is much different (i.e. system closer,
parallax higher). The problematic derivation of parallax from the Hipparcos data for the visual
binaries was discussed elsewhere (e.g. S{\"o}derhjelm 1999), but we cannot believe that the
difference is so large.

\begin{table}
 \begin{minipage}{130mm}
 \centering
  \caption{The orbital parameters of V949 Cen.}  \label{AstrOrbitV949Cen}
  \begin{tabular}{@{}c c c@{}}
\hline
Parameter       &  Value  &  Error  \\
 \hline
 $p_3$ [yr]     & 855.3   & 36.0    \\
 $T_0$ [yr]     & 1915.4  & 36.3    \\
 $e$            & 0.947   & 0.006   \\
 $a$ [arcsec]   & 2.463   & 0.012   \\
 $i$ [deg]      & 95.4    & 21.3    \\
 $\Omega$ [deg] & 114.8   & 47.6    \\
 $\omega$ [deg] & 106.6   & 17.5    \\
 \hline
\end{tabular}
\end{minipage}
\end{table}

\begin{figure}
  \centering
  \includegraphics[width=\textwidth]{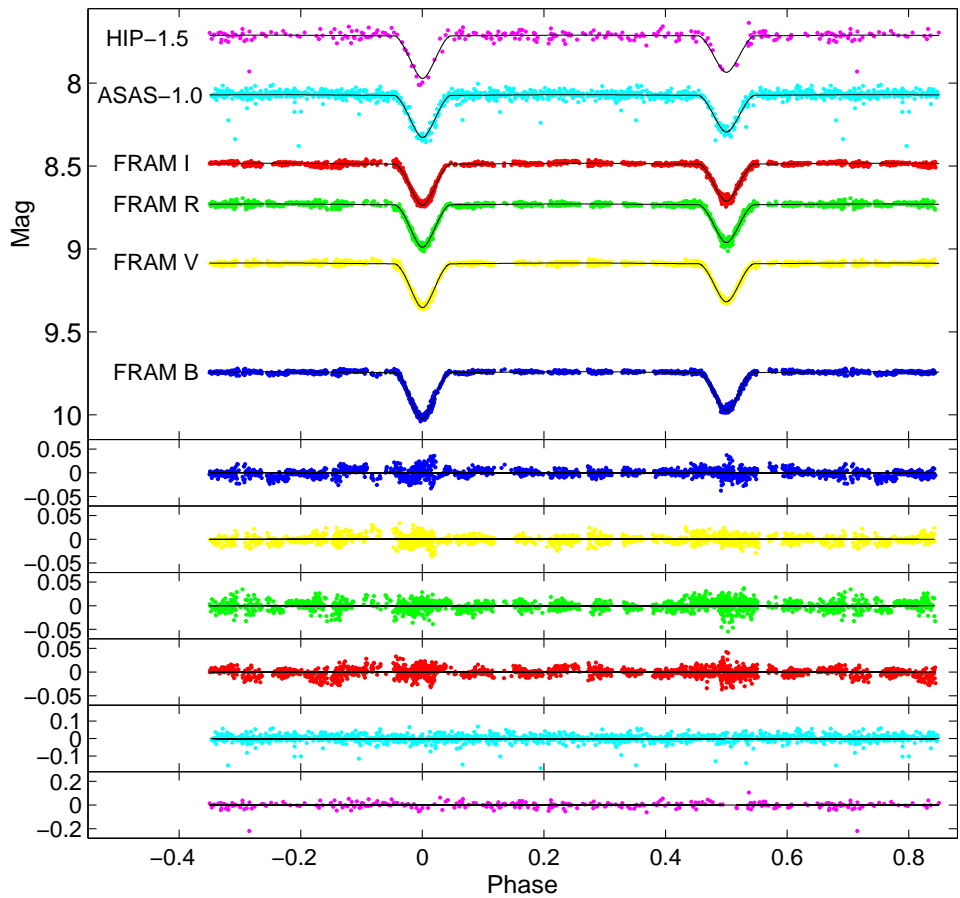}
  \caption{The available light curves of V949 Cen. The ASAS and Hipparcos curves were shifted
  in the y-axis for a better clarity. The bottom plots show the residuals of the fit.}
  \label{FigV949CenLC}
\end{figure}

\begin{figure}
  \centering
  \includegraphics[width=\textwidth]{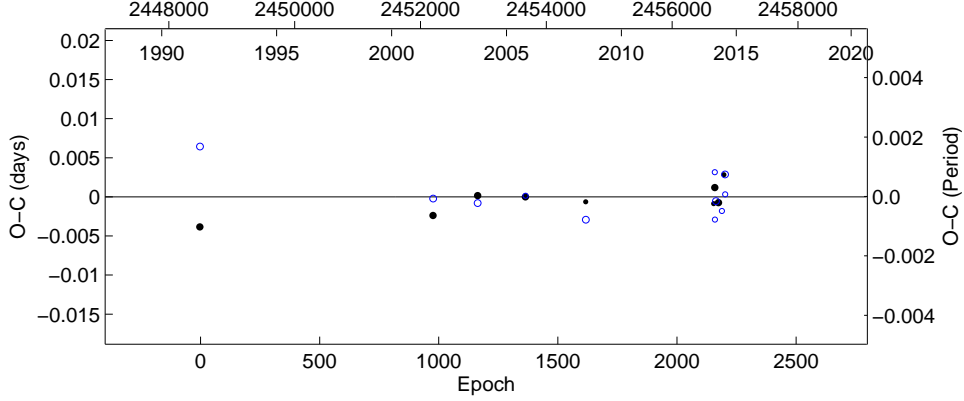}
  \caption{$O-C$ diagram of V949~Cen. Black symbols stand for the primary minima, the blue ones
  for the secondary minima. The larger the symbol, the higher the weight (higher the precision).}
  \label{FigV949CenOC}
\end{figure}

\begin{table*}
 \caption{The light-curve parameters of V949~Cen, as derived from our analysis.}
 \label{TableV949CenLC} \centering
\begin{tabular}{ c c | c c}
\hline
  Parameter     &  Value            & Parameter       &  Value            \\ \hline
 $HJD_0$        & 2448504.0866 $\pm$ 0.0096 & $L_{1,B}$ [\%] & 39.9 $\pm$ 1.4 \\
 $P$ [d]        & 3.7867969 $\pm$ 0.0000061 & $L_{2,B}$ [\%] & 30.9 $\pm$ 1.1 \\
 $i$ [deg]      & 83.66 $\pm$ 0.32  &  $L_{3,B}$ [\%] & 29.2 $\pm$ 0.9   \\
 $T_1$ [K]      & 5900 (fixed)      &  $L_{1,V}$ [\%] & 38.8 $\pm$ 0.9   \\
 $T_2$ [K]      & 5736 $\pm$ 42     &  $L_{2,V}$ [\%] & 31.5 $\pm$ 0.7   \\
  $\Omega_1$    & 7.429 $\pm$ 0.221 &  $L_{3,V}$ [\%] & 29.7 $\pm$ 0.5   \\
  $\Omega_2$    & 7.679 $\pm$ 0.204 &  $L_{1,R}$ [\%] & 38.1 $\pm$ 0.9   \\
 $F_1 = F_2$    & 0 (fixed)         &  $L_{2,R}$ [\%] & 31.6 $\pm$ 0.8   \\
 $A_1 = A_2$    & 0.5 (fixed)       &  $L_{3,R}$ [\%] & 30.3 $\pm$ 0.5   \\
 $G_1 = G_2$    & 0.32 (fixed)      &  $L_{1,I}$ [\%] & 37.3 $\pm$ 0.7   \\
 $q=M_2/M_1$    & 1.0 (fixed)       &  $L_{2,I}$ [\%] & 31.5 $\pm$ 0.5   \\
 $R_1/a$        & 0.156 $\pm$ 0.005 &  $L_{3,I}$ [\%] & 31.2 $\pm$ 0.5   \\
 $R_2/a$        & 0.150 $\pm$ 0.005 & $L_{1,ASAS}$ [\%] & 38.7 $\pm$ 0.7 \\
                &                   & $L_{2,ASAS}$ [\%] & 31.6 $\pm$ 0.6 \\
                &                   & $L_{3,ASAS}$ [\%] & 29.7 $\pm$ 0.6 \\
                &                   &  $L_{1,HIP}$ [\%] & 39.1 $\pm$ 1.7 \\
                &                   &  $L_{2,HIP}$ [\%] & 31.4 $\pm$ 0.9 \\
                &                   &  $L_{3,HIP}$ [\%] & 29.5 $\pm$ 2.1 \\  \hline
\end{tabular}
\end{table*}

Concerning the photometry of V949 Cen, we collected all available photometry from the Hipparcos
(Perryman et al. 1997) and ASAS (Pojma\'nski 2002) data together with our new observations from the
FRAM telescopes in $BVRI$ filters. 
There is clearly seen that the system is a detached one of Algol type, having two rather similar
components. The simultaneous analysis of all available photometry was carried out with the {\sc
PHOEBE} code, resulting in light curve parameters given in Table \ref{TableV949CenLC} and the light
curve fits presented in Figure \ref{FigV949CenLC}. As one can see, the value of the third light
resulted in a non-negligible contribution, indicating that the third component is probably very
similar to the eclipsing ones. This is in good agreement with the magnitude difference as presented
in already published papers (e.g. the Hipparcos catalogue presented the magnitude difference about
0.82~mag).

The ASAS, Hipparcos and FRAM photometry were also used to derive the times of minima for V949~Cen.
For computing the times of minima a classical Kwee-van Woerden (1956) method was used together with
our new AFP method (see Zasche et al. 2014). However, the twenty minima times plotted in the $O-C$
diagram (see Fig. \ref{FigV949CenOC}) show no period variation and the ephemerides suitable for any
future observations are given in Table \ref{TableV949CenLC}. Also from this plot it is obvious that
the innner orbit of the eclipsing pair is perfectly circular. All of the data points used for the
analysis are stored in the online-only Appendix tables.

\subsection{V358~Pup}

V358~Pup (also HD 51994, HIP 33487) is an Algol-type eclipsing binary discovered by the Hipparcos
satellite (Perryman et al. 1997). Its combined spectral type was derived as G5V (Houk 1978). It is
also known as a visual double and only eight observations were obtained since its discovery in
1926. However, no visible motion on the sky is presented, hence its period is probably rather long.
The latest observation of the double from 1993 indicates the separation of about 1$^{\prime\prime}$
at a position angle of 149$^\circ$. The only detailed study of the star is that one by Clausen et
al. (2001). However, their $uvby$ light curves were not analysed, only a brief statement about a
large fraction of the third light and several times of minima (showing mildly eccentric orbit) were
published.

\begin{figure}
  \centering
  \includegraphics[width=\textwidth]{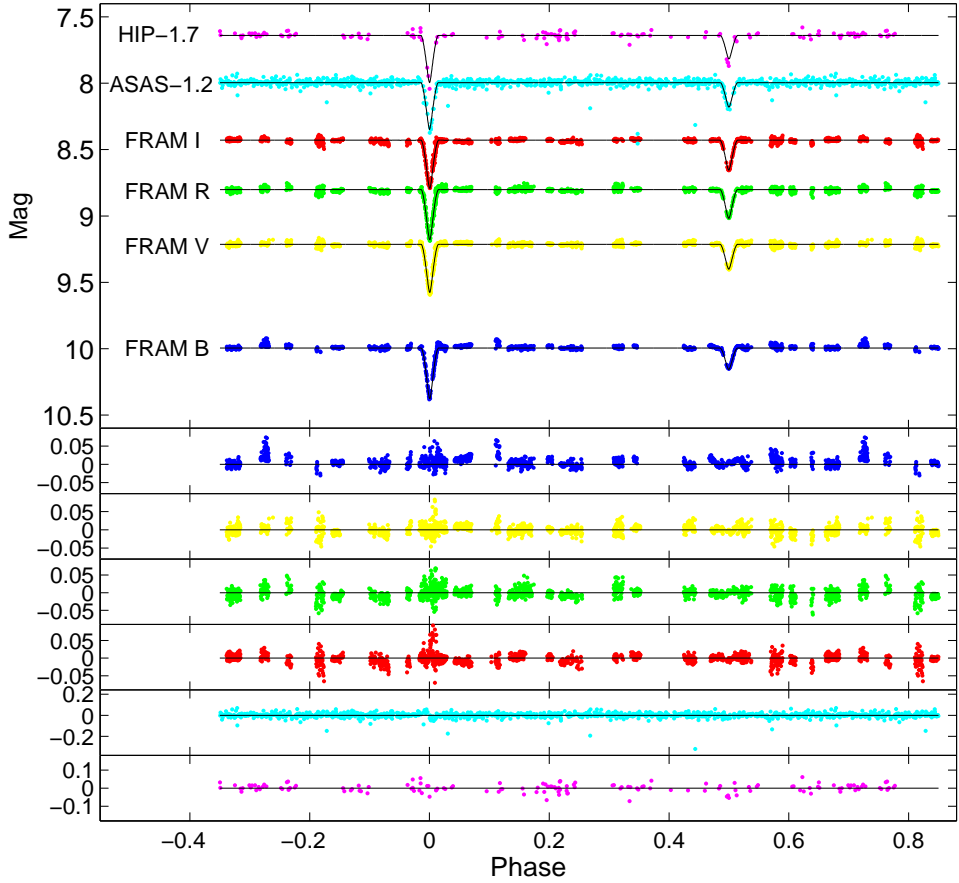}
  \caption{Light curves of V358 Pup. The ASAS and Hipparcos curves were shifted
  in the y-axis for a better clarity. The bottom plots show the residuals of the fit.}
  \label{FigV358PupLC}
\end{figure}

Our new FRAM light curves were analysed simultaneously with the other available photometry using
the {\sc PHOEBE} code. The results are plotted in Fig. \ref{FigV358PupLC}, while the parameters are
given in Table \ref{TableV358PupLC}. As one can see, the amount of the third light is really large,
which is in agreement with the published values of the magnitude difference between the two visual
components (the third component is only about 0.1 mag fainter than the combined magnitude of the
eclipsing pair).

\begin{table*}
 \caption{The light-curve parameters of V358~Pup, as derived from our analysis.}
 \label{TableV358PupLC} \centering
\begin{tabular}{ c c | c c}
\hline \hline
  Parameter     &  Value           &  Parameter       &  Value            \\ \hline
 $T_1$ [K]      & 5780 (fixed)     &   $i$ [deg]      & 89.35 $\pm$ 0.17  \\
 $T_2$ [K]      & 5157 $\pm$ 53    &   $HJD_0$        & 2450818.7131 $\pm$ 0.0009 \\
 $\Omega_1$     & 22.11 $\pm$ 0.50 &    $P$ [d]        & 6.7939106 $\pm$ 0.0000019 \\
 $\Omega_2$     & 20.17 $\pm$ 0.43 &    $e$            & 0.008 $\pm$ 0.005 \\
 $L_{1,B}$ [\%] & 33.8 $\pm$ 1.6   &   $\omega_0$ [deg]  & 97.4 $\pm$ 3.9 \\
 $L_{2,B}$ [\%] & 15.2 $\pm$ 1.4   &  $\mathrm{d}\omega / \mathrm{d}t$ [deg/cycle] & 0.002 $\pm$ 0.001 \\
 $L_{3,B}$ [\%] & 51.0 $\pm$ 0.5   & $F_1 = F_2$      & 0 (fixed)         \\
 $L_{1,V}$ [\%] & 32.9 $\pm$ 1.0   & $A_1 = A_2$      & 0.5 (fixed)       \\
 $L_{2,V}$ [\%] & 17.7 $\pm$ 1.1   & $G_1 = G_2$      & 0.32 (fixed)      \\
 $L_{3,V}$ [\%] & 49.4 $\pm$ 0.5   & $A$ [R$_\odot$]  & 18.67 $\pm$ 0.05  \\
 $L_{1,R}$ [\%] & 34.3 $\pm$ 1.4   & $q=M_2/M_1$      & 0.88  $\pm$ 0.05  \\
 $L_{2,R}$ [\%] & 20.4 $\pm$ 0.9   & $v_\gamma$ [km/s]& -2.23 $\pm$ 0.03  \\  \cline{3-4}
 $L_{3,R}$ [\%] & 45.2 $\pm$ 0.6   & \multicolumn{2}{c}{Derived physical quantities:} \\
 $L_{1,I}$ [\%] & 34.5 $\pm$ 0.9   & $M_1$ [M$_\odot$] & 1.01 $\pm$ 0.04 \\
 $L_{2,I}$ [\%] & 21.3 $\pm$ 1.1   & $M_2$ [M$_\odot$] & 0.89 $\pm$ 0.03 \\
 $L_{3,I}$ [\%] & 45.2 $\pm$ 0.8   & $R_1$ [R$_\odot$] & 0.88 $\pm$ 0.03 \\
 $L_{1,ASAS}$ [\%]& 33.6 $\pm$ 2.7 & $R_2$ [R$_\odot$] & 0.86 $\pm$ 0.02 \\
 $L_{2,ASAS}$ [\%]& 20.1 $\pm$ 2.6 & $K_1$ [km/s]      & 65.2 $\pm$ 2.6 \\
 $L_{3,ASAS}$ [\%]& 46.3 $\pm$ 4.0 & $K_2$ [km/s]      & 73.9 $\pm$ 2.7 \\
 $L_{1,HIP}$ [\%] & 32.1 $\pm$ 1.1 & $M_{bol,1}$ [mag] & 5.02 $\pm$ 0.08 \\
 $L_{2,HIP}$ [\%] & 17.9 $\pm$ 1.2 & $M_{bol,2}$ [mag] & 5.56 $\pm$ 0.06 \\
 $L_{3,HIP}$ [\%] & 50.0 $\pm$ 3.3 &                   &                 \\
  \hline
\end{tabular}
\end{table*}

\begin{table}
 \begin{minipage}{130mm}
 \centering
  \caption{The individual radial velocities of V358 Pup.}  \label{TabV358PupRVs}
  \begin{tabular}{@{}c c c c@{}}
\hline
                &  $RV_1$  &  $RV_2$  \\
 HJD            &  [km/s]  &  [km/s]  \\
 \hline
  2454889.75029 & -65.33 $\pm$ 0.47 &  69.54 $\pm$ 2.48 \\
  2454890.72131 & -51.67 $\pm$ 0.30 &  53.90 $\pm$ 0.91 \\
  2455118.89667 &  18.27 $\pm$ 0.44 & -26.06 $\pm$ 0.77 \\
  2455145.81532 &  32.13 $\pm$ 0.19 & -40.86 $\pm$ 0.37 \\
  2455469.82594 &  35.89 $\pm$ 1.17 & -46.11 $\pm$ 2.81 \\
  2455469.88360 &  38.19 $\pm$ 0.49 & -46.99 $\pm$ 1.51 \\
  2455479.74664 & -26.06 $\pm$ 0.27 &  24.21 $\pm$ 0.48 \\
  2455504.87144 &  61.51 $\pm$ 0.21 & -74.38 $\pm$ 0.34 \\
  2456214.88501 & -66.73 $\pm$ 0.47 &  70.46 $\pm$ 1.42 \\
 \hline
\end{tabular}
\end{minipage}
\end{table}

\begin{figure}
  \centering
  \includegraphics[width=\textwidth]{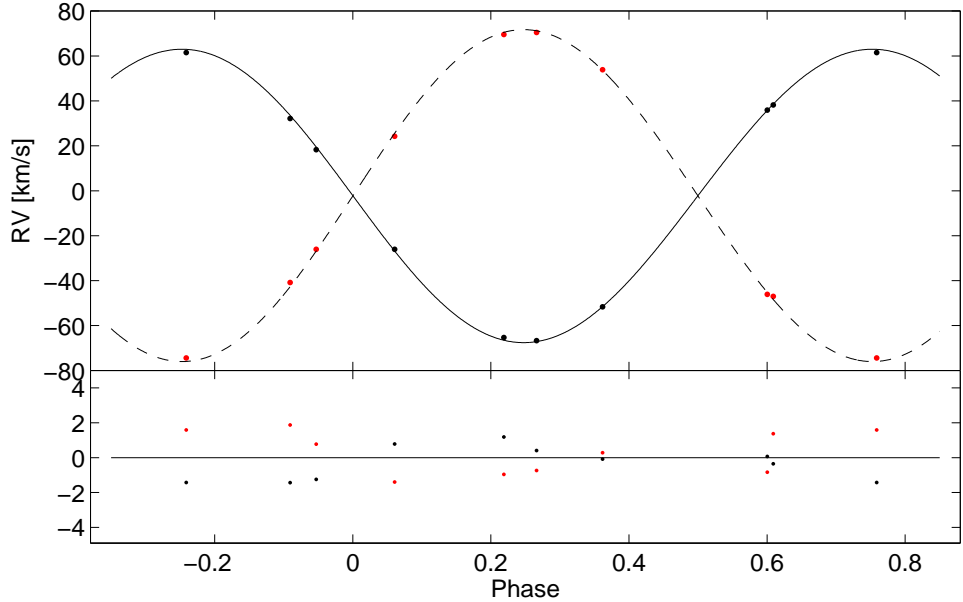}
  \caption{The radial velocities of V358 Pup, the black dots and the solid line stand
  for the primary, while the red dots and the dashed line stand for the secondary.}
  \label{FigV358PupRV}
\end{figure}

\begin{figure}
  \centering
  \includegraphics[width=\textwidth]{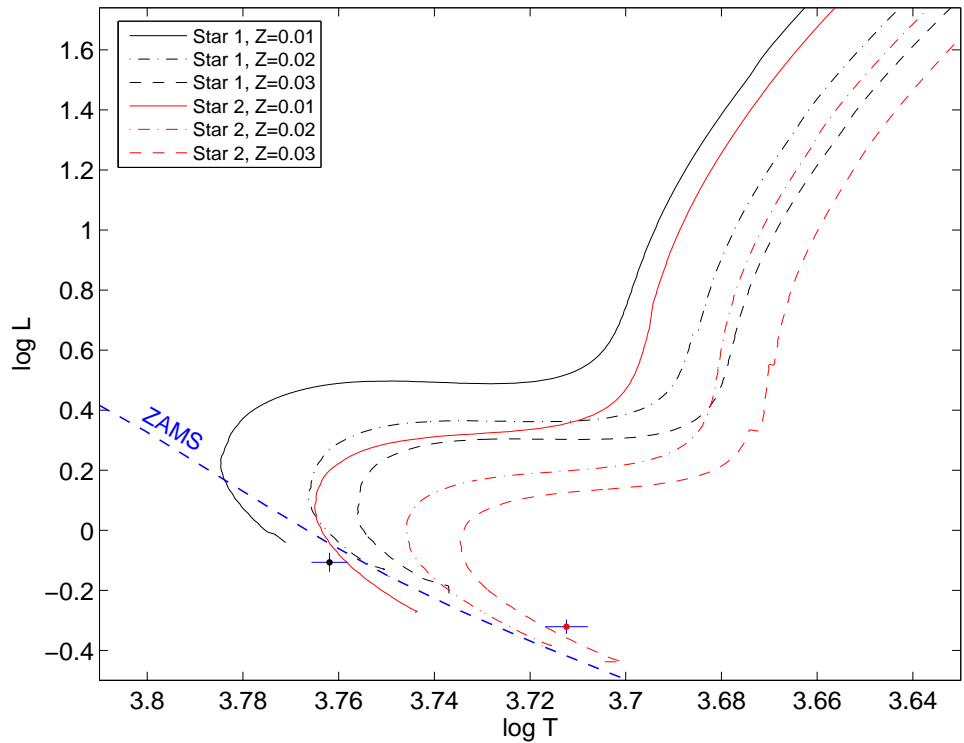}
  \caption{Evolutionary tracks for the system V358~Pup and both its components, star 1
  (primary in black) and star 2 (secondary in red) plotted for three different metallicities.
  The blue dashed line represents the ZAMS line for $Z=0.02$.}
  \label{FigV358PupHR}
\end{figure}

\begin{figure}
  \centering
  \includegraphics[width=\textwidth]{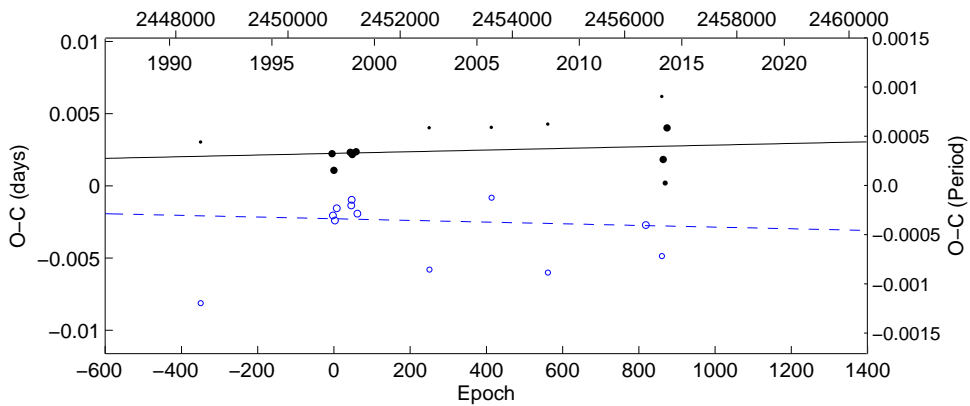}
  \caption{$O-C$ diagram of V358 Pup, showing very slow apsidal motion. Black symbols stand
  for the primary minima, the blue ones for the secondary minima, and the larger symbols
  correspond to the data points with the higher weights.}
  \label{FigV358PupOC}
\end{figure}

For the whole analysis we also used the ESO spectra downloaded from the online ESO archive
(altogether there were used nine HARPS spectra for the analysis\footnote{HARPS is a spectrograph
mounted on the 3.6-m ESO telescope at the La Silla Observatory. Its spectral range is from 3781 to
6911 A and the resolving power is 115 000.}).
 The original data were reduced using the standard ESO routines. The final radial
velocities (hereafter RV) used for the analysis were derived via a manual cross-correlation
technique (i.e. the direct and flipped profile of the spectral lines manually shifted on the
computer screen to achieve the best match) using program SPEFO (Horn et al. 1996, \v{S}koda 1996)
on several absorbtion lines in the measured spectral region (usually $Fe$ and $Ca$ lines). The
derived radial velocities are given in Table \ref{TabV358PupRVs}. There were also visible some
lines from the third body in the spectra, but all these lines show similar radial velocities, of
about -2.7~km$\cdot$s$^{-1}$ with no indication of change during the observed period (of about
3.6~yr).

As one can see from our plot given in Fig. \ref{FigV358PupRV}, the final RV curve is still not
perfect (as it is obvious from the residuals). We suppose that this effect can be partly caused by
the program {\sc PHOEBE} itself (and the relative weightening between the photometry and
spectroscopy) and partly also due to its rather uncertain eccentricity. From the pure RVs it would
indicate that the orbit has lower eccentricity, but from the photometry and the minima times there
resulted that the eccentricity should be higher.

As one can see from the final parameters given in Table \ref{TableV358PupLC}, the errors of masses
and radii are very low and one has to doubt about their values. With only a few data points and our
poor fit one cannot probably derive a mass with a precision of about 3-4\%. Therefore, we would
like to point out that the presented solution provided with the {\sc PHOEBE} program is a formal
one, based on the abovementioned assumptions (the inclusion of some future knowledge of these stars
can significantly shift our solution). Moreover, also the errors as resulted from the code are
purely mathematical ones and usually are strongly underestimated in the {\sc PHOEBE} program (Pr{\v
s}a \& Zwitter 2005, or PHOEBE
manual\footnote{http://phoebe-project.org/1.0/docs/phoebe$\_$manual.pdf}).

Another indication that the presented errors of parameters as given in Table \ref{TableV358PupLC}
are underestimated is a plot in Fig. \ref{FigV358PupHR}. The evolutionary tracks are plotted
together with the position of the two eclipsing components as resulted from our analysis (values of
temperatures and bolometric magnitudes taken from Table \ref{TableV358PupLC} with the error of
$T_1$ equal to error of $T_2$) and the theoretical curves are taken from the
EZ-web\footnote{http://www.astro.wisc.edu/$\sim$townsend/static.php?ref=ez-web} which is using a
well-known MESA code (Paxton et al. 2011). To conclude, here we deal very probably with two stars
still located on the main sequence or its vicinity and the models indicate that the stars have a
near solar metallicity.

From the available photometry also the minima times were derived. We collected the most recent FRAM
photometry together with the already published minima from the study by Clausen et al. (2001).
Moreover, also the older Hipparcos (Perryman et al. 1997) and ASAS (Pojma\'nski 2002) data were
used. All of these photometric data yielded 26 minima times, however some of them (especially the
ones from the Hipparcos and ASAS data) are of poor quality only. The resulting $O-C$ diagram is
plotted in Figure \ref{FigV358PupOC}. The ephemerides given in Table \ref{TableV358PupLC} refer to
the $y=0$ line in Fig. \ref{FigV358PupOC}. We analysed these data using the apsidal motion
hypothesis as described e.g. by Gim\'enez \& Garc\'{\i}a-Pelayo (1983). This led to the finding
that the apsidal motion is very slow, having the period of thousands of years and our present
analysis is still only very preliminary yet (and as one can see, the errors of apsidal motion
parameters are still very large). From our apsidal motion analysis there resulted that the
relativistic contribution to the total apsidal motion rate is rather small, being only about 11~\%,
and the value of $\log k_2 = -0.165$. This value can be compared with the evolutionary tracks as
published e.g. by Claret (2004), resulting in a finding that such a $\log k_2$ value can be
achieved with a 1~M$_\odot$ star with an age of about $2.5 \cdot 10^9$~yr.

\subsection{V1055~Sco}

The system V1055~Sco (also HD 148121, HIP 80603) is the only one W UMa-type binary in our sample of
stars. It has the orbital period of about 0.36~days only and rather shallow amplitude of
photometric variation of about 0.25~mag. It was discovered from the Hipparcos data (Perryman et al.
1997). On the other hand, Szczygie{\l} et al. (2008) included the star into their compilation of
the ASAS eclipsing binaries with the coronal activity. Torres et al. (2006) presented the spectral
type of V1055~Sco as G2V. The first astrometric observation of the visual double was carried out in
1925, published by Finsen (1941). Since then 13 positional measurements were obtained, but the
movement on the orbit is very slow (orbital period of the order of thousands of years probably).
The latest observation of the pair from 2010 indicates that the separation is of about
0.125$^{\prime\prime}$ at a position angle of 217$^\circ$. Magnitude difference between the two
visual components was derived to be from 0.0 to 0.3~mag (slightly different in different pass
bands).

\begin{table*}
 \caption{The light-curve parameters of V1055 Sco, as derived from our analysis.}
 \label{TableV1055ScoLC} \centering
\begin{tabular}{ c c c c c }
\hline \hline
 Photometry     &     FRAM             &     ASAS          &   SuperWASP       &     Hip           \\ \hline
  Parameter     &    Value             &     Value         &     Value         &     Value         \\ \hline
  $HJD_0$       & \multicolumn{4}{c}{2452050.7441 $\pm$ 0.0043} \\
  $P$ [d]       & \multicolumn{4}{c}{0.3636835 $\pm$ 0.0000038} \\
 $T_1$ [K]      &   \multicolumn{4}{c}{5850 (fixed)}  \\
 $T_2$ [K]      &   5751 $\pm$ 30      & 5632 $\pm$ 26     & 5881 $\pm$ 23     & 5499 $\pm$ 75     \\
 $i$ [deg]      &   89.4  $\pm$ 0.8    & 88.2 $\pm$ 1.3    & 88.6 $\pm$ 0.8    & 87.0 $\pm$ 1.9    \\
 $\Omega_1$     &   6.391 $\pm$ 0.019  & 6.746 $\pm$ 0.032 & 5.917 $\pm$ 0.021 & 5.407 $\pm$ 0.105 \\
 $\Omega_2$     &   6.293              & 6.422             & 5.782             & 4.410             \\
 $q=M_2/M_1$    &   3.0 $\pm$ 0.1      & 3.1 $\pm$ 0.1     & 2.7 $\pm$ 0.1     & 1.7 $\pm$ 0.1     \\
 $L_1$ [\%]     &  13.5 $\pm$ 0.2 (B)  & 12.8 $\pm$ 0.5    & 12.5 $\pm$ 0.3    & 14.3 $\pm$ 0.4    \\
 $L_1$ [\%]     &  13.2 $\pm$ 0.2 (V)  &                   &                   &                   \\
 $L_1$ [\%]     &  13.0 $\pm$ 0.2 (R)  &                   &                   &                   \\
 $L_1$ [\%]     &  12.9 $\pm$ 0.2 (I)  &                   &                   &                   \\
 $L_2$ [\%]     &  37.1 $\pm$ 0.2 (B)  & 37.8 $\pm$ 0.5    & 39.3 $\pm$ 0.4    & 35.0 $\pm$ 0.8    \\
 $L_2$ [\%]     &  37.3 $\pm$ 0.2 (V)  &                   &                   &                   \\
 $L_2$ [\%]     &  37.4 $\pm$ 0.2 (R)  &                   &                   &                   \\
 $L_2$ [\%]     &  37.4 $\pm$ 0.2 (I)  &                   &                   &                   \\
 $L_3$ [\%]     &  49.4 $\pm$ 0.3 (B)  & 49.4 $\pm$ 0.7    & 48.3 $\pm$ 0.6    & 50.7 $\pm$ 0.9    \\
 $L_3$ [\%]     &  49.5 $\pm$ 0.3 (V)  &                   &                   &                   \\
 $L_3$ [\%]     &  49.6 $\pm$ 0.2 (R)  &                   &                   &                   \\
 $L_3$ [\%]     &  49.7 $\pm$ 0.2 (I)  &                   &                   &                   \\
 $F_1 = F_2$    & \multicolumn{4}{c}{0.0 (fixed)}  \\
 $A_1 = A_2$    & \multicolumn{4}{c}{0.0 (fixed)}  \\
 $G_1 = G_2$    & \multicolumn{4}{c}{1.0 (fixed)}  \\
 $R_1/a$        & 0.297 $\pm$ 0.022    & 0.277 $\pm$ 0.017 & 0.303 $\pm$ 0.012 & 0.263 $\pm$ 0.019 \\
 $R_2/a$        & 0.517 $\pm$ 0.045    & 0.517 $\pm$ 0.035 & 0.502 $\pm$ 0.029 & 0.452 $\pm$ 0.040 \\
  \hline
\end{tabular}
\end{table*}


The light curve solution was a bit more complicated here than for the previous two systems. At
first, even the ephemerides were not very certain due to the fact that in our new FRAM photometry
both eclipses seem to be of similar depths (i.e. both temperatures probably very similar), but the
Hipparcos data show much larger difference between the minima depths. Therefore, we decided to use
the deeper Hipparcos minimum as the primary one and set the linear ephemerides according to this.
For the analysis and subsequent derivation of minima times for the period analysis we collected the
Hipparcos, ASAS, SuperWASP (Pollacco et al. 2006), Pi of the sky (Burd et al. 2005), OMC (Mas-Hesse
et al. 2004), and our new FRAM data. From these different data sets the ASAS, SuperWASP, and
Hipparcos seem to be the best ones for an analysis, however the total eclipses are only hardly
visible.

We solved the individual light curves of V1055~Sco separately. This is due to the fact that the
system undergoes some changes of the orbital period between the different datasets (see below) and
for the individual data the different ephemerides were used (fixed from our solution from the $O-C$
diagram below). The other two systems (V949~Cen and V358~Pup) were analysed with the complete
photometry simultaneously, because it is much more robust. For the result see Table
\ref{TableV1055ScoLC} with the final parameters of our fit. As one can see from the table, the
results are rather inconsistent with each other. The detached or contact configuration of the
binary did not led to an acceptable result. We applied a classical "q-search method" for deriving
the mass ratio of the binary (this can be applied because we deal with a semidetached system with
secondary filling its respective Roche lobe). The parameters as derived from some of the data sets
are too different from the others (e.g. the temperatures), that one has to doubt about
conclusiveness of the results. Generally, one can conclude that the role of primary and secondary
is very probably interchanged (due to mass ratio and luminosity), but the more massive component is
probably the cooler one. Also rather remarkable is quite high contribution of the third light in
the system. We can speculate about the origin of such different results based on different data
sets. One possible explanation is that the system is active and the shape of the light curve varies
in time. A role of some spots on the surface is also not excluded (slight asymmetry of the curve).
What is also remarkable is the fact that the values of inclination seem to be increasing in time,
but only a dedicated future observation can prove or rule out this hypothesis.

\begin{figure}
  \centering
  \includegraphics[width=\textwidth]{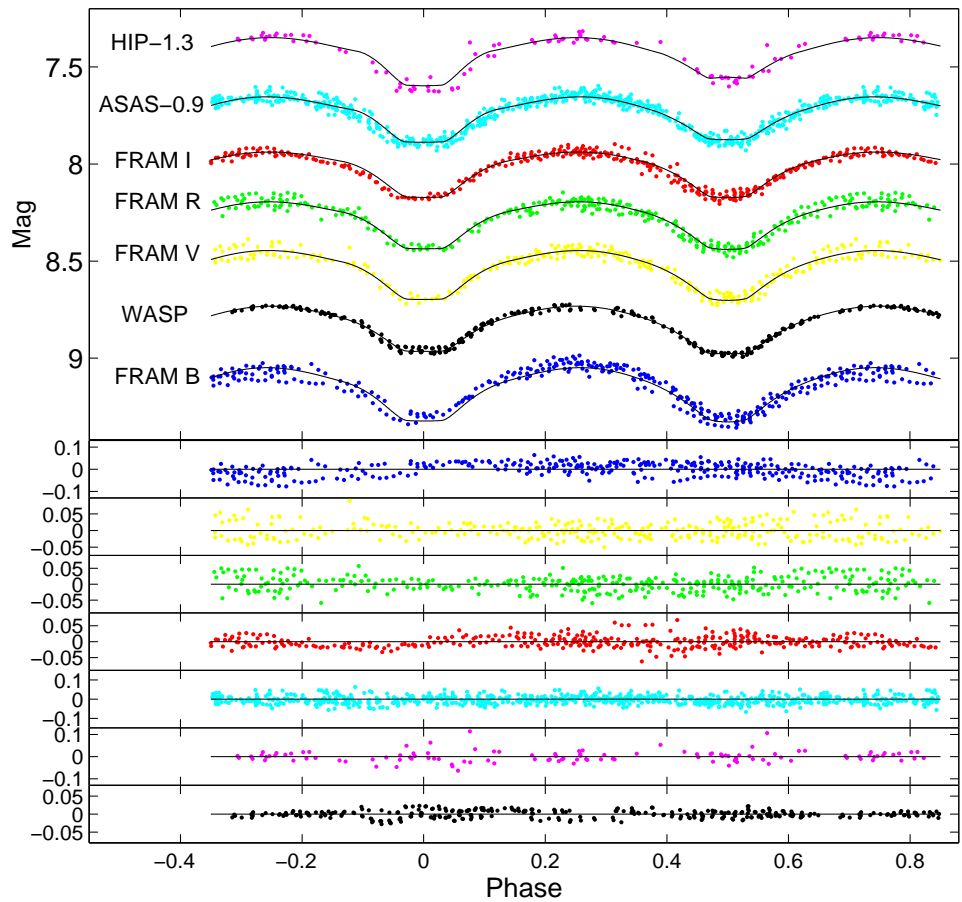}
  \caption{Light curves of V1055 Sco. The ASAS and Hipparcos curves were shifted
  in the y-axis for a better clarity. The bottom plots show the residuals of the fit.}
  \label{FigV1055ScoLC}
\end{figure}

\begin{table}
 \begin{minipage}{130mm}
 \centering
  \caption{The LITE orbital parameters of V1055 Sco.}  \label{LITEtabV1055Sco}
  \begin{tabular}{@{}c c c@{}}
\hline
Parameter     &  Value  &  Error \\
 \hline
 $p_3$ [yr]   & 36.7    & 12.5   \\
 $A$ [day]    & 0.031   & 0.009  \\
 $e$          & 0.00 (fixed)     \\
 $f(m_3)$ [M$_\odot$] & 0.118 & 0.037 \\
 \hline
\end{tabular}
\end{minipage}
\end{table}

\begin{figure}
  \centering
  \includegraphics[width=\textwidth]{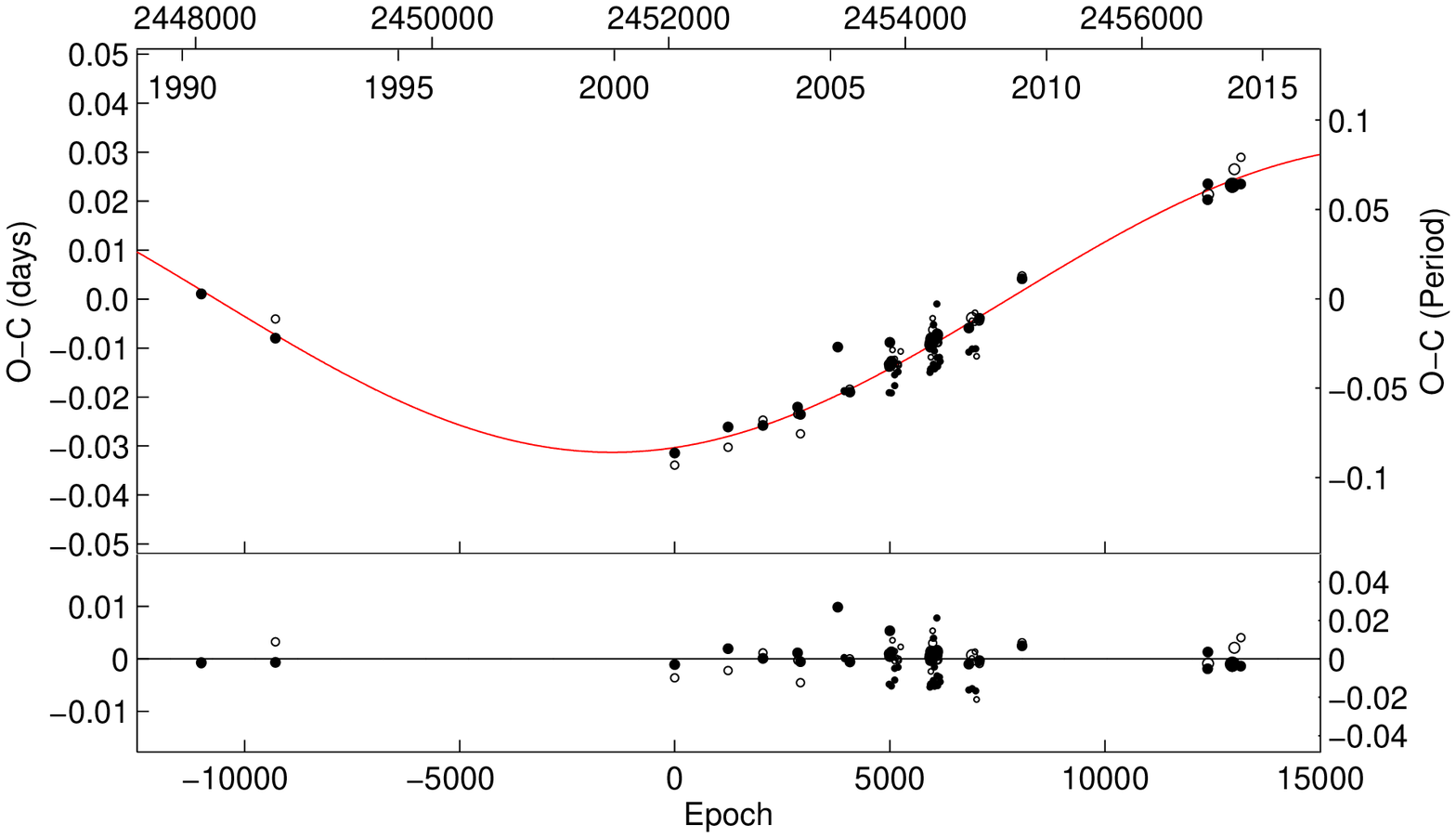}
  \caption{The $O-C$ diagram of V1055 Sco, plotted with all available times of minima observations. The dots stand
  for the primary, while the open circles for the secondary minima. The larger symbols stand for the observations
  with the higher precision. The solid curve represents the final LITE fit. The bottom plot shows the
  residuals of the fit.}
  \label{FigV1055ScoOC}
\end{figure}

From the available photometry, we also derived the minima times for a period analysis. Plotting the
$O-C$ diagram we found that there is a weak sinusoidal variation. Therefore, we applied a
hypothesis of the third body modulating the orbital period of the eclipsing pair, a so-called
Light-Time Effect (hereafter LITE), see e.g. Mayer (1990) or Irwin (1959). The result is given in
Table \ref{LITEtabV1055Sco} and the final fit is presented in Fig. \ref{FigV1055ScoOC}. The same as
for V358~Pup applies here and the ephemerides given in Table \ref{TableV1055ScoLC} represent the
linear ephemerides in Fig. \ref{FigV1055ScoOC}. As one can see, the coverage of the orbit is still
rather poor, hence we used only a circular orbit as a fit. If we collect more observations in the
upcoming years, some more detailed analysis would be possible. The errors of the parameters are so
large due to insufficient number of data points, especially close to the maximum in the $O-C$
diagram. Hence, the amplitude as well as the orbital period can only be estimated. We also tried to
identify this period variation with the visual component observed via interferometry, but this
attempt was not successful at all. Much different periods and eccentricities are needed for both
orbits to adequately fit the data. This would indicate that here we probably deal with a quadruple
stellar system. Some future observations in upcoming years would help us to identify the period
variation with the visual component or to prove its quadruple nature.

\section{Discussion and Conclusions}

We performed the first analysis of the light curves for three southern eclipsing binaries
containing the third components. However, these distant components have rather long orbital
periods, the only exception is V949~Cen, where we presented a preliminary orbital solution of the
visual double. All of these systems were observed remotely with the FRAM telescopes located in
Argentina and the light curve analysis revealed some interesting results. From the light curve fits
we also tried to derive the individual colors for the components and from the photometric indices
$B-V$ derive the temperatures. Such values can be compared with the temperatures as resulted from
the light curve fits to confirm whether the results are self-consistent. We found that the method
is not sensitive enough and its results are influenced by relatively large errors. However, the
resulted temperatures agree with the ones from the light curve fits given in Tables
\ref{TableV949CenLC}, \ref{TableV358PupLC}, and \ref{TableV1055ScoLC} within 100-150~K.

Probably the most problematic issue of our presented study is the fact that for the whole analysis
we assumed the value of primary temperatures for the light curve solution based only on the
photometric indices or the published spectral types. However, this can be quite problematic due to
the fact that for V358~Pup and V1055~Sco the most luminous components seem to be their tertiaries.

V949~Cen and V358~Pup were found to be well-detached systems, while the system V1055~Sco is
probably a semi-detached one. V358~Pup is showing slow apsidal motion of the inner orbit, but its
period is very long. For V1055~Sco we presented a preliminary solution for the period variation,
which can possibly be confirmed via a dedicated observation via spectroscopy and/or interferometry.

\Acknow{The operation of the robotic telescope FRAM is supported by the EU grant GLORIA (No. 283783
in FP7-Capacities program) and by the grant of the Ministry of Education of the Czech Republic
(MSMT-CR LG13007). This research has made use of the Washington Double Star Catalog maintained at
the U.S. Naval Observatory.  This investigation was supported by the Czech Science Foundation
grants no. P209/10/0715 and GA15-02112S. We also do thank the {\sc ASAS}, {\sc SuperWASP}, {\sc
OMC}, and {\sc Pi of the sky} teams for making all of the observations easily public available.
This research has made use of the SIMBAD and VIZIER databases, operated at CDS, Strasbourg, France
and of NASA's Astrophysics Data System Bibliographic Services.}

\section{Appendix Tables}

\newpage

\begin{table}
 \caption{Heliocentric minima times for the studied systems.}
 \label{TableMin} \centering 
  \scriptsize
\begin{tabular}{ c c c c | c c c c }
\hline \hline
    Star  &  HJD         & Error    &  Source                   &     Star  &  HJD        & Error    & Source    \\
          & 2400000+     & [days]   &                           &           & 2400000+    & [days]   &           \\ \hline
 V949 Cen & 48492.72269  & 0.00564  &  HIP                      & V1055 Sco & 53909.70012 & 0.00085  &   Pi of the sky \\
 V949 Cen & 48494.62638  & 0.00530  &  HIP                      & V1055 Sco & 53910.60608 & 0.00081  &   Pi of the sky \\
 V949 Cen & 52199.99788  & 0.00455  &  ASAS                     & V1055 Sco & 53938.61031 & 0.00141  &   Pi of the sky \\
 V949 Cen & 52201.89344  & 0.00458  &  ASAS                     & V1055 Sco & 53941.70325 & 0.00112  &   Pi of the sky \\
 V949 Cen & 52908.13137  & 0.00287  &  ASAS                     & V1055 Sco & 54257.02189 & 0.00084  &   ASAS          \\
 V949 Cen & 52910.02381  & 0.00170  &  ASAS                     & V1055 Sco & 54257.20413 & 0.00120  &   ASAS          \\
 V949 Cen & 53669.27729  & 0.00242  &  ASAS                     & V1055 Sco & 54626.89174 & 0.00121  &   ASAS          \\
 V949 Cen & 53671.17083  & 0.00712  &  ASAS                     & V1055 Sco & 54627.07413 & 0.00150  &   ASAS          \\
 V949 Cen & 54627.33616  & 0.00742  &  ASAS                     & V1055 Sco & 54986.40150 & 0.00080  &   ASAS          \\
 V949 Cen & 54629.22726  & 0.00306  &  ASAS                     & V1055 Sco & 54986.58393 & 0.00108  &   ASAS          \\
 V949 Cen & 56660.84562  & 0.00529  &  FRAM                     & V1055 Sco & 56554.62091 & 0.00481  &   FRAM          \\
 V949 Cen & 56677.89022  & 0.00629  &  FRAM                     & V1055 Sco & 56557.53361 & 0.00261  &   FRAM          \\
 V949 Cen & 56679.78165  & 0.00474  &  FRAM                     & V1055 Sco & 56559.53170 & 0.00300  &   FRAM          \\
 V949 Cen & 56681.67097  & 0.00867  &  FRAM                     & V1055 Sco & 56762.83229 & 0.00217  &   FRAM          \\
 V949 Cen & 56696.82051  & 0.00050  &  FRAM                     & V1055 Sco & 56763.74187 & 0.00219  &   FRAM          \\
 V949 Cen & 56736.58168  & 0.00297  &  FRAM                     & V1055 Sco & 56781.74747 & 0.00271  &   FRAM          \\
 V949 Cen & 56791.48915  & 0.01289  &  FRAM                     & V1055 Sco & 56835.75144 & 0.00502  &   FRAM          \\
 V949 Cen & 56823.68156  & 0.01061  &  FRAM                     & V1055 Sco & 56836.66608 & 0.00317  &   FRAM          \\
 V949 Cen & 56840.72217  & 0.00448  &  FRAM                     & V1055 Sco & 53862.41972 & 0.00219  &   SuperWASP     \\
 V949 Cen & 56844.50644  & 0.01036  &  FRAM                     & V1055 Sco & 53862.59620 & 0.00082  &   SuperWASP     \\
 V358 Pup & 48440.84746  & 0.00202  &  HIP                      & V1055 Sco & 53881.33135 & 0.00101  &   SuperWASP     \\
 V358 Pup & 48444.23326  & 0.00163  &  HIP                      & V1055 Sco & 53881.50766 & 0.00072  &   SuperWASP     \\
 V358 Pup & 52517.19479  & 0.00046  &  ASAS                     & V1055 Sco & 53891.51779 & 0.00656  &   SuperWASP     \\
 V358 Pup & 52520.58192  & 0.00071  &  ASAS                     & V1055 Sco & 53909.33457 & 0.00315  &   SuperWASP     \\
 V358 Pup & 53624.60224  & 0.00245  &  ASAS                     & V1055 Sco & 53921.33692 & 0.06034  &   SuperWASP     \\
 V358 Pup & 53627.99432  & 0.00671  &  ASAS                     & V1055 Sco & 53961.34467 & 0.00046  &   SuperWASP     \\
 V358 Pup & 54630.10122  & 0.00204  &  ASAS                     & V1055 Sco & 54207.56117 & 0.00161  &   SuperWASP     \\
 V358 Pup & 54633.48790  & 0.00248  &  ASAS                     & V1055 Sco & 54208.46336 & 0.00023  &   SuperWASP     \\
 V358 Pup & 50784.74575  & 0.00030  &  Clausen et al. (2001)    & V1055 Sco & 54211.56150 & 0.00068  &   SuperWASP     \\
 V358 Pup & 50801.72630  & 0.00020  &  Clausen et al. (2001)    & V1055 Sco & 54212.46443 & 0.00041  &   SuperWASP     \\
 V358 Pup & 50818.71420  & 0.00020  &  Clausen et al. (2001)    & V1055 Sco & 54212.65260 & 0.00105  &   SuperWASP     \\
 V358 Pup & 50835.69555  & 0.00020  &  Clausen et al. (2001)    & V1055 Sco & 54216.46515 & 0.00054  &   SuperWASP     \\
 V358 Pup & 50869.66590  & 0.00020  &  Clausen et al. (2001)    & V1055 Sco & 54216.64935 & 0.00240  &   SuperWASP     \\
 V358 Pup & 51110.85360  & 0.00100  &  Clausen et al. (2001)    & V1055 Sco & 54230.47724 & 0.00387  &   SuperWASP     \\
 V358 Pup & 51127.83470  & 0.00030  &  Clausen et al. (2001)    & V1055 Sco & 54231.56459 & 0.00383  &   SuperWASP     \\
 V358 Pup & 51134.62900  & 0.00100  &  Clausen et al. (2001)    & V1055 Sco & 54236.65049 & 0.00081  &   SuperWASP     \\
 V358 Pup & 51144.82295  & 0.00020  &  Clausen et al. (2001)    & V1055 Sco & 54237.38584 & 0.00056  &   SuperWASP     \\
 V358 Pup & 51151.61695  & 0.00020  &  Clausen et al. (2001)    & V1055 Sco & 54237.55969 & 0.00436  &   SuperWASP     \\
 V358 Pup & 51212.76235  & 0.00030  &  Clausen et al. (2001)    & V1055 Sco & 54238.47345 & 0.00231  &   SuperWASP     \\
 V358 Pup & 51236.53670  & 0.00100  &  Clausen et al. (2001)    & V1055 Sco & 54239.38623 & 0.00183  &   SuperWASP     \\
 V358 Pup & 56379.52621  & 0.00288  &  FRAM                     & V1055 Sco & 54244.29166 & 0.00010  &   SuperWASP     \\
 V358 Pup & 56661.48240  & 0.00482  &  FRAM                     & V1055 Sco & 54246.29080 & 0.00137  &   SuperWASP     \\
 V358 Pup & 56664.86830  & 0.00153  &  FRAM                     & V1055 Sco & 54246.47502 & 0.00750  &   SuperWASP     \\
 V358 Pup & 56688.65368  & 0.01859  &  FRAM                     & V1055 Sco & 54247.37822 & 0.00218  &   SuperWASP     \\
 V358 Pup & 56722.62160  & 0.00345  &  FRAM                     & V1055 Sco & 54247.56558 & 0.00169  &   SuperWASP     \\
 V358 Pup & 56756.59497  & 0.05620  &  FRAM                     & V1055 Sco & 54248.29360 & 0.00082  &   SuperWASP     \\
V1055 Sco & 48048.77191  & 0.00165  &  HIP                      & V1055 Sco & 54266.30300 & 0.00151  &   SuperWASP     \\
V1055 Sco & 48048.95356  & 0.00095  &  HIP                      & V1055 Sco & 54268.29609 & 0.00450  &   SuperWASP     \\
V1055 Sco & 48674.66215  & 0.00694  &  HIP                      & V1055 Sco & 54268.47414 & 0.00071  &   SuperWASP     \\
V1055 Sco & 48674.84790  & 0.00250  &  HIP                      & V1055 Sco & 54271.20528 & 0.00030  &   SuperWASP     \\
V1055 Sco & 52050.71267  & 0.00081  &  ASAS                     & V1055 Sco & 54271.38183 & 0.00242  &   SuperWASP     \\
V1055 Sco & 52050.89202  & 0.00089  &  ASAS                     & V1055 Sco & 54271.56979 & 0.00004  &   SuperWASP     \\
V1055 Sco & 52501.32187  & 0.00199  &  ASAS                     & V1055 Sco & 54272.29748 & 0.00202  &   SuperWASP     \\
V1055 Sco & 52501.49955  & 0.00157  &  ASAS                     & V1055 Sco & 54272.47287 & 0.00102  &   SuperWASP     \\
V1055 Sco & 52795.90580  & 0.00071  &  ASAS                     & V1055 Sco & 54273.38677 & 0.00393  &   SuperWASP     \\
V1055 Sco & 52796.08872  & 0.00063  &  ASAS                     & V1055 Sco & 54274.29147 & 0.00205  &   SuperWASP     \\
V1055 Sco & 53090.1272   & 0.0025   &  Zasche et al. (2009)     & V1055 Sco & 54274.47862 & 0.00243  &   SuperWASP     \\
V1055 Sco & 53090.3076   & 0.0031   &  Zasche et al. (2009)     & V1055 Sco & 54288.47677 & 0.00439  &   SuperWASP     \\
V1055 Sco & 53113.04009  & 0.00149  &  ASAS                     & V1055 Sco & 54294.29489 & 0.00034  &   SuperWASP     \\
V1055 Sco & 53113.21796  & 0.00163  &  ASAS                     & V1055 Sco & 54537.60107 & 0.00159  &   SuperWASP     \\
V1055 Sco & 53429.09480  & 0.00140  &  OMC                      & V1055 Sco & 54562.51977 & 0.00294  &   SuperWASP     \\
V1055 Sco & 53483.6384   & 0.0013   &  Ogloza et al. (2008)     & V1055 Sco & 54564.51429 & 0.00065  &   SuperWASP     \\
V1055 Sco & 53483.8200   & 0.0005   &  Ogloza et al. (2008)     & V1055 Sco & 54586.52250 & 0.00185  &   SuperWASP     \\
V1055 Sco & 53530.37204  & 0.00091  &  ASAS                     & V1055 Sco & 54588.34293 & 0.00106  &   SuperWASP     \\
V1055 Sco & 53530.55329  & 0.00079  &  ASAS                     & V1055 Sco & 54594.33636 & 0.00549  &   SuperWASP     \\
V1055 Sco & 53869.88015  & 0.00288  &  ASAS                     & V1055 Sco & 54601.42667 & 0.00723  &   SuperWASP     \\
V1055 Sco & 53870.05697  & 0.00184  &  ASAS                     & V1055 Sco & 54614.52627 & 0.00086  &   SuperWASP     \\
V1055 Sco & 53909.51280  & 0.00155  &  Pi of the sky            &           &             &          &                 \\
 \hline
\end{tabular} 
\end{table}

\end{document}